\documentstyle[epsfig,aps,multicol]{revtex}

\title{Replica Theory of Granular Media}
\author{Jeferson J. Arenzon}  
\address{Instituto de F\'{\i}sica -- UFRGS \\
CP 15051 -- 91501-970 -- Porto Alegre RS -- BRAZIL \\
E-mail: {\tt arenzon@if.ufrgs.br} 
\\ Homepage: {\tt www.if.ufrgs.br/\char 126 arenzon}}
\date{June 25, 1998}

\begin{document}
\maketitle

\begin{abstract}
An infinite range spin glass like model for granular systems
is introduced and studied through the replica mean field formalism.
Equilibrium, density dependent properties under vibration and gravity
are obtained. 
\end{abstract}

\begin{multicols}{2}
\narrowtext

Handling of granular material is present in many agricultural and
industrial processes and several fundamental practical problems are
still unsolved.
Besides that, their unusual static and flow properties
\cite{science} offer a challenging problem from 
the theoretical point of view, and despite the huge effort that has
been devoted in recent years, are far from being fully understood.
Since thermal energy plays no role here, excitations can be
achieved by externally shaking or shearing the system, enabling them to wander 
through
the many microscopic configurations available for a fixed macroscopic density.
Under vibration, a multitude of fascinating phenomena show up,
like heap formation, convection cells, size or shape segregation, surface 
waves, etc (see \cite{science} for a review and references).
Another effect is the logarithmically slow rate of the
density increase as the system suffers a sequence of taps \cite{knight}.
This slow relaxation phenomenon under perturbations, signalling complex
cooperative movements of the particles, resembles the one found 
in systems with many metastable states like glasses and spin glasses.

The analogy between glassy and granular behavior has been suggested
some time ago \cite{science} and stressed recently along with
the role of geometric frustration \cite{ConiglioHerrmann}. This
frustration arises from the excluded volume of the grains, imposing
restrictions on their relative positions.
For glass forming liquids, a simple frustrated lattice gas (FLG) has 
been introduced \cite{mimmo1} that takes into account these
steric effects and bridges complex fluids (glasses)
and complex magnets (spin glasses). 
An infinite range version \cite{ans96,sna97} 
has also been studied in the framework of replica theory, 
yielding a very rich phase diagram. In order to apply this model
for granular systems, Nicodemi {\it et al} \cite{mimmo} 
introduced the effects of gravity in the model, studying a tilted 2$d$
lattice while applying a sequence of taps, the particles following a 
diffusion-like Monte Carlo dynamics. Among several interesting
properties, analogous to those found in real experiments, the system 
displays an inverse logarithmic compactation behavior,
reversible-irreversible cycles, aging, as well as a 
localization transition, signalled by a zero diffusion constant, in 
which the particles get trapped into dynamical local cages. This 
transition point 
seems to correspond to the Reynolds (or dilatancy) transition observed
in real systems. A lattice model has several advantages.
From the computational point of view, the simulation cost is much
lower than molecular dynamics. From the theoretical side, besides
the amenability to theoretical investigation explored here, it helps
in grasping the fundamental concepts involved responsible for the
complex observed phenomena.

\begin{figure}
\vspace{-1cm}
\centerline{
\setlength{\unitlength}{0.240900pt}
\ifx\plotpoint\undefined\newsavebox{\plotpoint}\fi
\sbox{\plotpoint}{\rule[-0.200pt]{0.400pt}{0.400pt}}%
\begin{picture}(1200,809)(0,0)
\font\gnuplot=cmr10 at 10pt
\gnuplot
\sbox{\plotpoint}{\rule[-0.200pt]{0.400pt}{0.400pt}}%
\put(93,162){\makebox(0,0)[l]{\ldots }}
\put(93,405){\makebox(0,0)[l]{\ldots }}
\put(93,647){\makebox(0,0)[l]{\ldots }}
\put(963,162){\makebox(0,0)[l]{\ldots }}
\put(963,405){\makebox(0,0)[l]{\ldots }}
\put(963,647){\makebox(0,0)[l]{\ldots }}
\put(1035,162){\makebox(0,0)[l]{$\ell$ + 1}}
\put(1035,405){\makebox(0,0)[l]{$\ell$}}
\put(1035,647){\makebox(0,0)[l]{$\ell$ - 1}}
\put(745,526){\makebox(0,0)[l]{$J_{ij}^{\ell-1}$}}
\put(818,283){\makebox(0,0)[l]{$J_{ij}^{\ell}$}}
\put(165,162){\circle*{18}}
\put(310,162){\circle*{18}}
\put(455,162){\circle*{18}}
\put(600,162){\circle*{18}}
\put(745,162){\circle*{18}}
\put(890,162){\circle*{18}}
\put(165,405){\circle*{18}}
\put(310,405){\circle*{18}}
\put(455,405){\circle*{18}}
\put(600,405){\circle*{18}}
\put(745,405){\circle*{18}}
\put(890,405){\circle*{18}}
\put(165,647){\circle*{18}}
\put(310,647){\circle*{18}}
\put(455,647){\circle*{18}}
\put(600,647){\circle*{18}}
\put(745,647){\circle*{18}}
\put(890,647){\circle*{18}}
\put(310,405){\usebox{\plotpoint}}
\multiput(308.92,405.00)(-0.499,0.835){287}{\rule{0.120pt}{0.768pt}}
\multiput(309.17,405.00)(-145.000,240.407){2}{\rule{0.400pt}{0.384pt}}
\multiput(165.58,643.81)(0.499,-0.835){287}{\rule{0.120pt}{0.768pt}}
\multiput(164.17,645.41)(145.000,-240.407){2}{\rule{0.400pt}{0.384pt}}
\put(310.0,405.0){\rule[-0.200pt]{0.400pt}{58.298pt}}
\put(310,405){\usebox{\plotpoint}}
\multiput(310.58,405.00)(0.499,0.835){287}{\rule{0.120pt}{0.768pt}}
\multiput(309.17,405.00)(145.000,240.407){2}{\rule{0.400pt}{0.384pt}}
\multiput(453.92,643.81)(-0.499,-0.835){287}{\rule{0.120pt}{0.768pt}}
\multiput(454.17,645.41)(-145.000,-240.407){2}{\rule{0.400pt}{0.384pt}}
\multiput(310.00,405.58)(0.599,0.500){481}{\rule{0.579pt}{0.120pt}}
\multiput(310.00,404.17)(288.798,242.000){2}{\rule{0.290pt}{0.400pt}}
\put(310,405){\usebox{\plotpoint}}
\multiput(310.00,405.58)(0.899,0.500){481}{\rule{0.819pt}{0.120pt}}
\multiput(310.00,404.17)(433.300,242.000){2}{\rule{0.410pt}{0.400pt}}
\multiput(741.60,645.92)(-0.899,-0.500){481}{\rule{0.819pt}{0.120pt}}
\multiput(743.30,646.17)(-433.300,-242.000){2}{\rule{0.410pt}{0.400pt}}
\multiput(310.00,405.58)(1.199,0.500){481}{\rule{1.059pt}{0.120pt}}
\multiput(310.00,404.17)(577.803,242.000){2}{\rule{0.529pt}{0.400pt}}
\put(600,405){\usebox{\plotpoint}}
\multiput(596.61,403.92)(-0.895,-0.500){483}{\rule{0.816pt}{0.120pt}}
\multiput(598.31,404.17)(-433.306,-243.000){2}{\rule{0.408pt}{0.400pt}}
\multiput(165.00,162.58)(0.895,0.500){483}{\rule{0.816pt}{0.120pt}}
\multiput(165.00,161.17)(433.306,243.000){2}{\rule{0.408pt}{0.400pt}}
\multiput(597.60,403.92)(-0.597,-0.500){483}{\rule{0.577pt}{0.120pt}}
\multiput(598.80,404.17)(-288.802,-243.000){2}{\rule{0.289pt}{0.400pt}}
\put(600,405){\usebox{\plotpoint}}
\multiput(598.92,401.80)(-0.499,-0.838){287}{\rule{0.120pt}{0.770pt}}
\multiput(599.17,403.40)(-145.000,-241.401){2}{\rule{0.400pt}{0.385pt}}
\multiput(455.58,162.00)(0.499,0.838){287}{\rule{0.120pt}{0.770pt}}
\multiput(454.17,162.00)(145.000,241.401){2}{\rule{0.400pt}{0.385pt}}
\put(600.0,162.0){\rule[-0.200pt]{0.400pt}{58.539pt}}
\put(600,405){\usebox{\plotpoint}}
\multiput(600.58,401.80)(0.499,-0.838){287}{\rule{0.120pt}{0.770pt}}
\multiput(599.17,403.40)(145.000,-241.401){2}{\rule{0.400pt}{0.385pt}}
\multiput(743.92,162.00)(-0.499,0.838){287}{\rule{0.120pt}{0.770pt}}
\multiput(744.17,162.00)(-145.000,241.401){2}{\rule{0.400pt}{0.385pt}}
\multiput(600.00,403.92)(0.597,-0.500){483}{\rule{0.577pt}{0.120pt}}
\multiput(600.00,404.17)(288.802,-243.000){2}{\rule{0.289pt}{0.400pt}}
\end{picture}
} 
\vspace{0.5cm}
\caption{System architecture with $L$ layers, 
connections being only between nearest neighbors layers.}
\label{model.fig}
\end{figure}
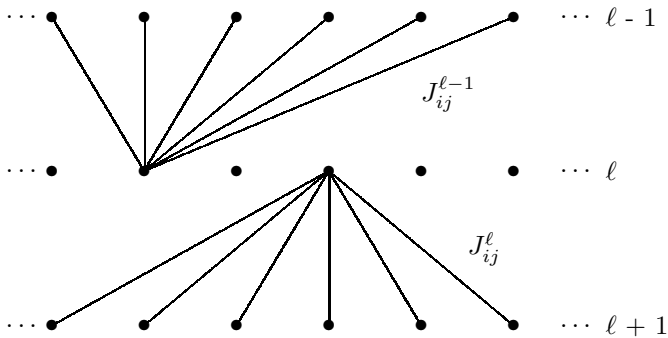

In the same spirit as Sherrington and Kirkpatrick \cite{SK}
introduced an exactly solvable version of the Edwards-Anderson
model \cite{EA},
here we introduce an infinite range version of the FLG considering
$L$ layers of $N$ sites, connections being only between nearest
neighbors layers (see fig.\ref{model.fig}). Each site may be occupied by a
particle ($n_i=0,1$) having, for simplicity, only two possible
spatial orientation, $S_i=\pm 1$. Although the grains can assume several
spatial orientations (usually infinite), here we take the simplest case.
The steric effects imposed by neighboring particles are felt as
restrictions on the particle orientation, what is included in the
hamiltonian as quenched, gaussian distributed lattice bonds $J_{ij}^{\ell}$
with $\overline{J_{ij}^{\ell}}=0$ and $\overline{(J_{ij}^{\ell})^2}=J/N$.  
Although the geometric
frustration on the internal degrees of freedom should be considered as 
annealed at low densities
and almost quenched at high ones, the quenched approximation is
sufficiently good as can be seen from the 2$d$ results \cite{fp,mimmo1}.
Each layer has its own chemical
potential satisfying $\mu_{\ell+1}>\mu_{\ell}$ (counting from top
to bottom) that accounts for the effect of gravity and 
$\mu_{\ell}=g\ell/L$ in order to have a constant force. Thus, we 
consider the following Hamiltonian:
\begin{equation}
{\cal H} = - \sum_{i<j} \sum_{\ell=1}^{L-1}
\left(J_{ij}^{\ell} S_i^{\ell}S_j^{\ell+1} + \frac{K}{N}\right)
n_i^{\ell}n_j^{\ell+1} 
-\sum_{\ell=1}^{L-1} \mu_{\ell} \sum_i n_i^{\ell} 
\label{hamiltonian}
\end{equation}
The parameter $K=-1+K'$ may tune the repulsive/attractive
($K'$ negative/positive) interaction between
particles \cite{sna97} and may be important to treat wet powders.
Here, as in the original model, we consider $K=-J=-1$ ($K'=0$).
Notice that the value $-1$ appears originally in order to
recover, in the limit $J\gg 1$, the Frustrated Percolation (FP) \cite{fp}
constraint of only allowing
non frustrated loops to be fully occupied. In this limit, two given neighboring
sites, only could be occupied ($n_in_j=1$) if the corresponding orientations
satisfied the local disorder, $J_{ij}S_iS_j=1$, otherwise at least one site
should be empty ($n_in_j=0$).
It is also
important to point out that by changing the value of $K$, several new 
qualitatively different phases do appear, in both frustrated 
or not versions of the Blume-Emery-Griffiths (BEG) model (see \cite{sna97} 
and references therein). 
In the limit where all sites are occupied, we get
a layered version of the SK model (L-SK). This model has
a continuous transition from a spin glass ($q_{\ell}\neq 0$) phase to a 
paramagnetic 
one ($q_{\ell}=0,\forall \ell$) at $T_c^{SK}=1/\sqrt{2x_c}$ where $x_c$ is the
lowest positive root of a polynomial recursively obtained by 
$P_L^{SK}(x)=P_{L-1}^{SK}(x)-x^2 P_{L-2}^{SK}(x)$ with 
$P_0^{SK}=P_1^{SK}=1$. A similar result \cite{BeBeIgPa}
has been found for a modified mean-field version of the McCoy-Wu model
\cite{McCoyWu}. This critical 
temperature approaches unity as $L\to\infty$.

In evaluating the free energy, the trace is restricted to states with a given 
density $\rho$. Using standard techniques for dealing with disordered
systems \cite{MPV}  and assuming replica symmetry, the free energy reads
\begin{eqnarray}
f &=& \frac{\beta}{4} \sum_{\ell} q_{\ell} q_{\ell+1} 
   - \frac{1}{2}\left( \frac{\beta}{2}-1\right)\sum_{\ell} 
     d_{\ell} d_{\ell+1}+ \beta \sum_{\ell} t_{\ell} d_{\ell} \nonumber \\
  &&  -\sum_{\ell}(\mu_0+\mu_{\ell}) d_{\ell} 
     -\frac{\beta}{2} \sum_{\ell} r_{\ell} q_{\ell}  
     -\frac{L}{\beta}\ln 2 
     + \mu_0 \rho L \nonumber\\ 
  && -\frac{1}{\beta} \sum_{\ell}\int{\cal D}z\:\ln\left\{ 1 +
     \cosh\left( \beta \sqrt{r_{\ell}} z \right)
     \mbox{e}^{-\Xi_{\ell}}    \right\}
\end{eqnarray}
where ${\cal D}z=dz/\sqrt{2\pi}\exp(-z^2/2)$ and the temperature
$T=\beta^{-1}$ is a measure of the vibration imposed to the system.
The order parameters are a diluted Edwards-Anderson 
$q_{ab}^{\ell} = \left\langle S^{a\ell} n^{a\ell} 
S^{b\ell} n^{b\ell}\right\rangle$ and the density 
$d_a^{\ell} = \left\langle n^{a\ell} \right\rangle$
while $\mu_0$ accounts
for the constraint $\rho = L^{-1} \sum_{\ell} d_{\ell}$ and
\begin{eqnarray}
\Xi_{\ell} &=& \frac{\beta^2}{4} (q_{\ell+1} + q_{\ell-1}) 
               - \frac{\beta}{2} \left( \frac{\beta}{2} -1\right)
                    (d_{\ell+1} + d_{\ell-1}) \nonumber \\
          &&  - \beta (\mu_{\ell}+\mu_0)
\end{eqnarray}
The saddle point equations are $r_{\ell} = (q_{\ell-1} + q_{\ell+1})/2$ and
\begin{eqnarray}
q_{\ell} &=& \int{\cal D}z\frac{\sinh^2 (\beta
\sqrt{r_{\ell}}z)}{\left[
           \mbox{e}^{\Xi_{\ell}}+
           \cosh(\beta \sqrt{r_{\ell}}z)\right]^2} \\
d_{\ell} &=& \int{\cal D}z\frac{\cosh (\beta 
           \sqrt{r_{\ell}}z)}{\mbox{e}^{\Xi_{\ell}}+
           \cosh(\beta  \sqrt{r_{\ell}}z)}
\end{eqnarray}
where  $q_0=d_0=q_{L+1}=d_{L+1}=0$.
A global order parameter may be introduced as $Q=L^{-1}\sum_{\ell} q_{\ell}$.

\begin{figure}
\centerline{\epsfig{file=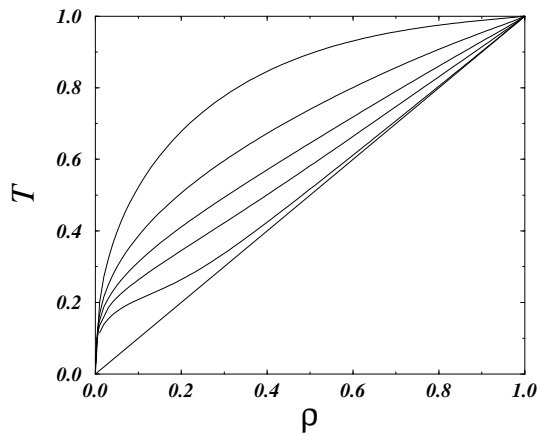,width=6cm,angle=270}}
\caption{Phase diagram $T$ versus $\rho$ for $L=100$ and $g$ from
0 (straight line) to 5 (top curve), showing the disordered phase 
($q_{\ell}=0, \forall \ell$) and a spin glass phase 
($q_{\ell}\neq 0$).}
\label{tcl}
\end{figure}

There is a critical temperature $T_c$ above which all $q_{\ell}$ are zero,
that is, $Q=0$. Figure \ref{tcl} shows, for $L=100$, the critical
temperature as a function of density for
several values of $g$. Notice that the bigger the density, the stronger
should be the vibration in order to get a fluid state, similar
to what happens for fixed density and increasing gravity.
When $g=0$ (no gravity), $d_{\ell}=\rho \;(\forall \ell)$ and
$T_c=\rho T_c^{SK}$
where $T_c^{SK}$ is the critical temperature of the $L$-layered
SK model (see below).
On the other limit, when $g\to\infty$ (strong gravity), in analogy with the 
layered SK model, the critical temperature is related with the 
smallest positive root $x^*$ of a given polynomial, $T_c = 1/\sqrt{2x^*}$,
depending on the density
$\rho$. For $0\leq\rho\leq 1/L$, we have $T_c=0$ because all particles 
occupy the lowest layer and do not interact. For $1/L \leq \rho \leq 2/L$, 
some sites occupy the second lowest layer and the critical temperature is 
obtained with the root of $P_2(x)$.
 In general, for $(n-1)/L \leq\rho\leq n/L$,
the relevant polynomial is $P_n(x)$. These polynomials are
obtained  recursively by
\begin{equation}
P_{\ell}(x)=\frac{P_{\ell-1}(x)}{\delta_{\ell L}(\rho L-L+1)^2
+(1-\delta_{\ell L})} - x^2 P_{\ell-2}(x)
\end{equation}
with $P_0(x)=P_1(x)=1$. In this limit, the critical temperature goes to unity
for large $L$.

\begin{figure}
\centerline{\epsfig{file=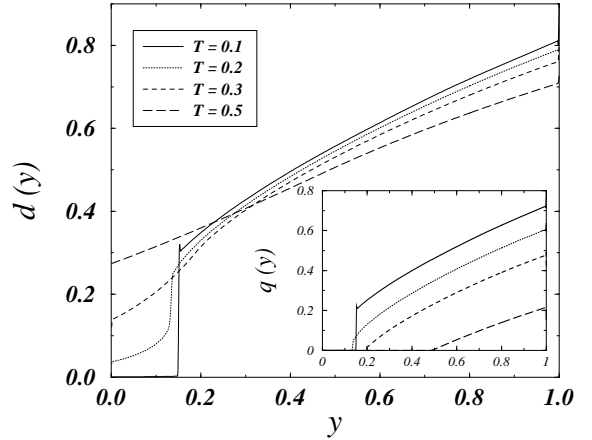,width=6cm,angle=270}}
\caption{Density profile as a function of $y=\ell/L$ for $L=1000$,
$\rho=0.5$, $g=1$ and several
values of $T$. At high temperatures we recover the original mean
field FLG as the density becomes uniform. Inset: profile of $q_{\ell}$. 
Although the curve seems to go to zero above
a certain layer, it actually does not do so for finite $L$. The onset
of fluidization arrives at once for every layer.}
\label{dprofile}
\end{figure}

\begin{figure}
\centerline{\epsfig{file=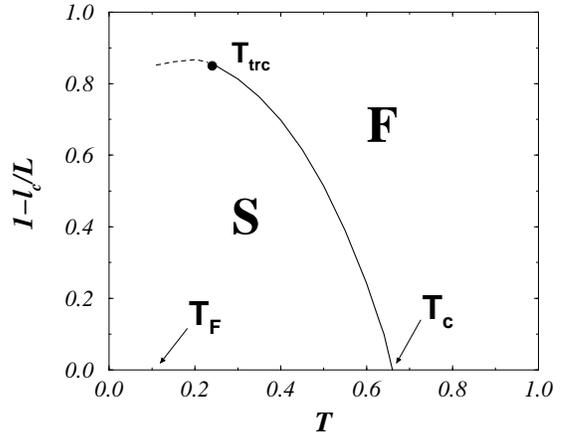,width=6cm,angle=270}}
\caption{Transition line $1-\ell_c/L$ showing the higher, fluidized layers and
the lower, solid ($q_{\ell}\neq 0$) ones for $g=1$, $\rho=0.5$ and large
$L$. The onset of fluidization occurs at
$T_F\simeq 0.11$ while the system is completely fluidized above $T_c\simeq
0.66$. For $T<T_F$ we have $q_{\ell}\neq 0$ for every non empty layer and
since there is no longer a solid-fluid interface, just the material
surface, we do not show the line.
The transition is first order for $T_F<T<T_{trc}\simeq 0.24$ and
continuous up to $T_c$.}
\label{lc}
\end{figure}

The density and $q$ profiles are shown in fig. \ref{dprofile} as a
function of the system height $\ell/L$. Interestingly enough, the density
profile shows,
even at low temperature, that frustration effects are important in
preventing a close packed configuration, signaled by a density
lower than one.
Its important to stress that, in analogy with the FLG \cite{ans96},
there are two regimes of densities depending on the gravity: one is the
L-SK regime for large $g$
with the particles settling in the lowest possible layers disregard
the geometric effects and the second and most interesting one, shown here,
where the
steric effects become important. Although the $q$-profile
seems to vanish above a given height, it actually does not do so
for finite $L$ and as we approach the continuous limit ($L\to\infty$),
a new transition settles down. From experiments and
molecular dynamics simulation \cite{Jason,Taguchi},
as a function of the vibration, the
system passes from a solid-like behavior to a fluid-solid coexistence
region. After this onset of fluidization, the top layers become
fluid-like, the particles having a great mobility. The interface between those
regions decreases its height as the vibration increases.
Although in this context a fluid regime means a situation where
the particles present translational mobility, here we consider
the case where the mobility is orientational, that is, a fluid
layer will be one having $q_{\ell}=0$. However, in some
sense, both are a measure of the amount of geometrical constraints
imposed on the particles by their neighbors. Moreover, when simulating
the 3$d$  system in absence of gravity \cite{mimmo1}, it can be seen that
at the glass transition, the
diffusion coefficient does vanish, corroborating our use of the term.
For increasing values of $L$ we can extrapolate the numerical
results and find
the critical layer, $\ell_c(T)$, separating the fluid region
($q_{\ell}=0$) and the solid one ($q_{\ell}\neq 0$). The temperature
where this first happens, signalling the onset of fluidization, is
denoted by $T_F$ and the transition is discontinuous from $T_F$
up to $T_{trc}$, while continuous for $T_{trc}<T<T_c$.
Notice that below $T_F$ all non-empty layers have $q_{\ell}\neq 0$.  
The point $T_{trc}$ is reminiscent
of the tricritical point found in the disordered BEG model (remember that
we have a varying chemical potential in the vertical axis).
These information is summarized in figure \ref{lc}.
Notice that although in the figure \ref{dprofile} the
top most layers have $q_{\ell}=0$ below $T_F$, they are empty. Moreover,
the density profile
has been measured both in real experiments \cite{CleRaj}
as well as in simulations, showing that in the steady state, independently
of the phase of the up and down motion of the heap,
the density profile is always preserved. This fact, indicating that
configurational properties may be obtained through appropriate averages,
independently of the dynamics, supports our equilibrium results
and has also led to thermodynamic theories of granular systems \cite{Hong}.
Also, as the granular temperature increases, there is an elevation of
the center of mass of the system \cite{LuHeBl}, given by
$h_{CM} = (\rho L)^{-1} \sum_{\ell} \ell d_{\ell}$
and this is a function of both temperature and gravity. 

In conclusion, we introduced an infinite range version of a
frustrated lattice gas model \cite{ConiglioHerrmann,mimmo} for 
granular systems and applied, to our knowledge for the first time, 
the replica formalism to these systems. 
In this mean field version, we are able to study stationary
properties, obtaining the vibration, density and gravity dependent
phase digram as well as information on the density
profile and the onset of fluidization.

There is a multitude of possibilities that still remain to be explored. 
First of all, the
stability of the replica symmetric solution employed here and the behavior
of the response functions, like the compressibility, for example.
Also, the effects
of segregation on polydisperse mixtures observed experimentally
\cite{science} and in the 2$d$ model\cite{CaCoHeLoNi}, can be included
by allowing different degrees of frustration for each type of particle
in the Hamiltonian \cite{AreNi}. Results on the 2$d$ model show that
the out of equilibrium dynamics, when the system is subject to small, continuous
shaking \cite{NiCo98}, present aging in the two times correlation function
$C(t,t_w)$ for the bulk density, remanescent of the glassy nature of the
model. In view of this, it would be extremely interesting to study in detail
the mean-field equations for the dynamics \cite{LeticiaJorge}.
Different coupling distributions, e.g. bimodal,
may be studied by means of the TAP formalism \cite{tap}, and
in the limit of large $L$, the density and $q$ profile should obey a set
of coupled differential equations, from which some analytical
results may be obtained.
From the
simulational point of view, research is in course to study how compactation
and segregation are affected by interpolating from the 2$d$ model to the mean
field case, along with the effects of attraction ($K>-1$) between the
particles, as in wet powders. Besides that, the fluidization transition 
may be detected on the simulation by measuring the mean squared displacement
for each particle, $R^2_i(t)$ (instead of the system average) and comparing
with its mean height.

\noindent
{\bf Acknowledgments:} I thank A. Coniglio, N. Lemke, M. Nicodemi, L. Peliti 
and M. Sellitto for very fruitful discussions, 
and the warm hospitality of the Dipartimento di Scienze Fisiche
(Universit\`a di Napoli, Italy) during my stay.
Work partially supported by the Brazilian agencies CNPq and FINEP.

\end{multicols}
\end{document}